\documentclass[fleqn,twoside]{article}
\usepackage{fleqn,espcrc2,epsf}

\usepackage{graphicx}

\newcommand{\be}{\begin{equation}}
\newcommand{\ee}{\end{equation}}
\newcommand{\AmS}{{\protect\the\textfont2
  A\kern-.1667em\lower.5ex\hbox{M}\kern-.125emS}}

\newcommand{\mres}{m_{\rm res}}
\title{Domain-Wall Fermions at Strong Coupling
\thanks{This work was conducted on the QCDSP machines at Columbia
     University and the RIKEN-BNL Research Center. LL and RM are
     supported by the US DOE.
}}
\author{L.~Levkova\address{Department of Physics, Columbia 
University, New York, NY, 10027} and R.~Mawhinney$^{\rm a}$}
\begin{document}
\bibliographystyle{apsrev}

\begin{abstract}
The DWF formulation becomes increasingly
problematic at gauge couplings for which $a^{-1}<2$ GeV, where the roughness
of the gauge field leads to increased explicit chiral symmetry
breaking ($\mres$). This problem becomes especially severe for 
sufficiently strong coupling where the underlying 4-dimensional 
Wilson theory is in the Aoki phase. We review our attempts to find a suitable 
modification of the gauge and/or the fermion action 
which would allow the DWF method to work reliably
at stronger coupling.
\vspace{1pc}
\end{abstract}
\maketitle
\section{INTRODUCTION}
Domain-wall fermions have the advantage of having 
an exact chiral symmetry in the limit of an infinite 5th dimension on the lattice, even 
at a non-zero lattice spacing.
In combination with the DBW2 gauge action \cite{dbw2} in quenched simulations 
it has been shown that the $\mres$ is about two orders of magnitude
smaller than its value for the Wilson gauge action \cite{rbc}. This effect has been attributed to the 
properties of DBW2 to suppress lattice dislocations,
which contribute directly to the $\mres$, 
and to make perturbative corrections small \cite{rbc,inst}. However, in full QCD calculations 
at coarse lattice spacings,
for the combination of DWF/DBW2, the $\mres$
is still not satisfactory small.
The quark-gluon system is in the Aoki phase where the DWF formulation 
suffers from the violations of chiral symmetry and locality.
Our goal is to find a suitable modification of the gauge action and/or the fermion 
action which would allow us to conduct a thermodynamics calculation with DWF.
%
\section{EFFECTS OF THE ADDITION OF AN ADJOINT TERM TO DBW2}
Our studies of the plaquette distributions of dynamical DWF/DBW2 configurations \cite{me}
showed that with the increase of the coupling the tails of the distributions 
extend to more negative plaquette values, which are associated with
the appearance of greater number of lattice dislocations and increased $\mres$.
These findings gave us the idea to modify
the DBW2 action by adding an adjoint term to it with the expectation that the modified action will 
suppress more strongly the negative tails of the plaquette
distributions and thus it will get rid of the lattice dislocations.
The modified DBW2 action has the following form:
\begin{small}
\begin{eqnarray}
\label{eq:asym_gauge_action}
S_{\small \rm  dbw2+adj}\!\!\! 
  &=&\!\!\! -\frac{\beta}{N_c}\left((1-8c_1)\left[c_f
     \sum_{x;\mu<\nu} {\rm Re} {\rm Tr} P_{\mu\nu}(x)\right.\right. \nonumber\\
 &&\hspace{-1.9cm}+ \left.\left.\frac{c_a}{N_c}\sum_{x;\mu<\nu}\left|{\rm Tr} P_{\mu\nu}(x)\right|^2 \right]
 +  c_1 \sum_{x;\mu \neq \nu } {\rm Re} {\rm Tr}R_{\mu\nu}(x)\right),
     \nonumber
\end{eqnarray}
\end{small}
\hspace{-0.14cm}where we choose $c_f+2c_a=1$, to keep the normalization
for this three term action the same as for the plaquette action
in the continuum limit.
The tail shortening effect on the plaquette distributions is an expected result of the decreasing 
of the effective coupling for the fundamental plaquettes 
in the action, when the coefficient $c_a$ is negative.

\begin{figure}[ht]
\epsfxsize=\hsize
\begin{center}
\epsfbox{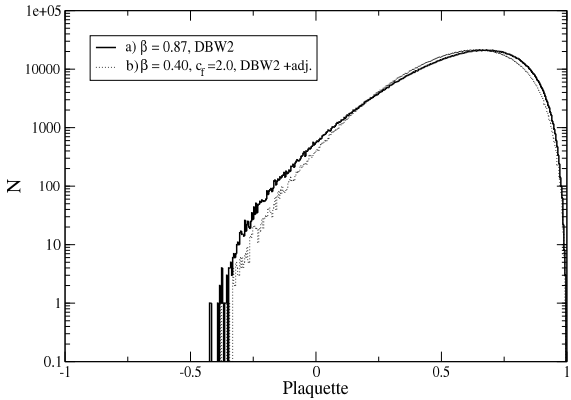}
\end{center}
\vspace{-1.4cm}
\caption{Plaquette distributions for quenched runs $a$ and $b$. Run $a$
has $\beta=0.87$ and $a^{-1}\approx 1.3$ GeV. Run $b$ has $\beta=0.40$, $c_f=2.0$ 
and $a^{-1}\approx 0.93$ GeV. Both runs have volume $16^3\times 32$ and $L_s=12$.}
\label{fig:quplaq}
\end{figure}

\begin{figure}
\epsfxsize=\hsize
\begin{center}
\epsfbox{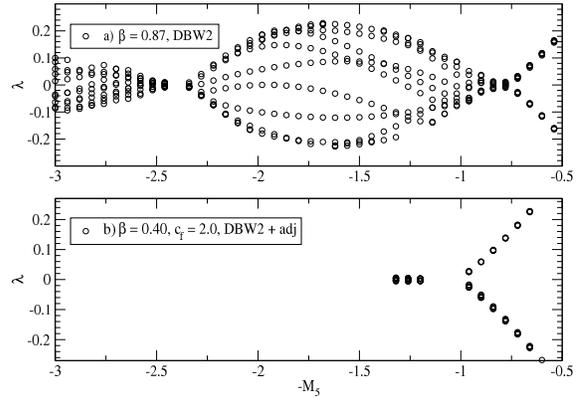}
\end{center}
\vspace{-1.4cm}
\caption{Spectral flow of $\gamma_5D_W$ for run $a$ and $b$ from Figure~\ref{fig:quplaq}.}
\label{fig:sflow_qu}
\end{figure}

In the quenched case, 
we chose to match the plaquette distributions of two runs, one with the modified 
DBW2 and the other with the original DBW2, and compare their $\mres$ and lattice spacing.
Figure~\ref{fig:quplaq} shows that although the plaquette distribution of the modified DBW2 run $b$ 
has a shorter negative tail, its $\mres=0.0156(2)$ is actually an order of magnitude 
larger than the corresponding value of $\mres=0.00125(3)$ for the unmodified DBW2 run $a$. Furthermore
run $a$ is out of the Aoki phase and run $b$ according to its closed spectral flow gap shown on 
on Figure~\ref{fig:sflow_qu}, is in that phase.\vspace{-0.4cm}
\vspace{-0.4cm}
In conclusion, we attempted to ``cure'' the Aoki phase by changing the action at the scale of the lattice spacing
by adding an adjoint term and shortening the tails of the plaquette distributions.
The negative tail of the plaquette distribution although probably contributing
to $\mres$ out of the Aoki phase, when the system is in that phase, does not play a significant role in 
determining $\mres$. The extent of the tail of the plaquette distribution into negative values 
is not a  primary reason for the onset of the Aoki phase and although the DBW2 with 
the added adjoint term does reduce it at a given lattice spacing, 
that does not help to push the system out of the Aoki phase.
\section{EFFECTS OF A LARGE RECTANGLE TERM IN THE GAUGE ACTION}
The rectangle term in the gauge action reduces the effective coupling for the  
plaquette term and ``smoothes'' the gauge field on the lattice by introducing 
some amount of non-locality at the scale of two lattice spacings. Since we have 
concluded from the previous section that changing the action at the
scale of one plaquette does not influence the Aoki phase, which is a more long-range global phenomenon,
we want to investigate the effects of adding more of the less-local rectangle term. In practice we change
the coefficient $c_1=-1.4069$ in the DBW2 action with a larger negative value.

\begin{figure}
\epsfxsize=\hsize
\begin{center}
\epsfbox{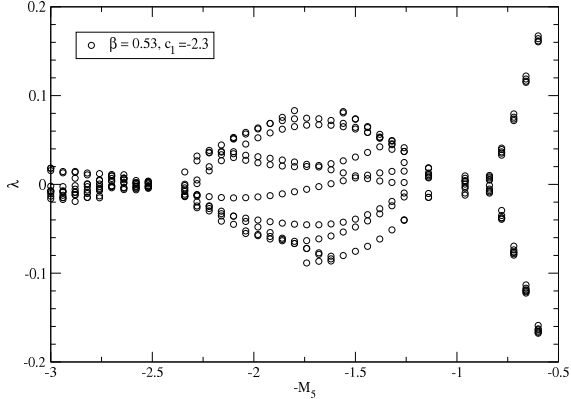}
\end{center}
\vspace{-1.4cm}
\caption{Spectral flow of $\gamma_5D_W$ for a quenched run with DBW2-style plaquette plus rectangle
action where $\beta=0.53$ and $c_1=-2.3$. Volume $16^3\times 32$ and $L_s=12$.}
\label{fig:qurect}
\end{figure}

We performed a quenched calculation with a DBW2-style action with $\beta=0.53$ and
$c_1=-2.3$, which has $a^{-1}\approx 1.0$ GeV
and $\mres =0.0035(1)$.
The spectral flow for this run on Figure~\ref{fig:qurect}, shows a gap and some crossings which
we believe means that this run is out of the Aoki phase even at that coarse $a$. 
However to use such a large rectangle term for physics calculations could have undesired 
consequences, since the short distance physics is distorted by the introduced non-locality.
\section{TWISTED MASS TERM IN THE DWF DIRAC OPERATOR}
As another attempt to solve the problem posed by the Aoki phase 
we would like to add an irrelevant to the physics term in the DWF action, which  
would regularize the approach to the Aoki phase. 
We want to experiment with the term 
$im_\tau\gamma_5\tau_3\delta_{x,x^\prime}\delta_{s,s^\prime}$,
called a ``twisted mass'' term. As an analogue of $M_5$, the twisted mass term provides
another parameter, $m_\tau$, which could be independently tuned to hopefully achieve 
a better localization of the bound to the domain-walls light chiral states, which, if possible,
could help with the Aoki phase problem. 
We have investigated the free case analytically 
to make sure that the twisted mass term addition does not break down 
the most important properties of the DWF formulation,
namely, the existence of light bound to the domain walls states. 
\begin{figure}
\epsfxsize=\hsize
\begin{center}
\epsfbox{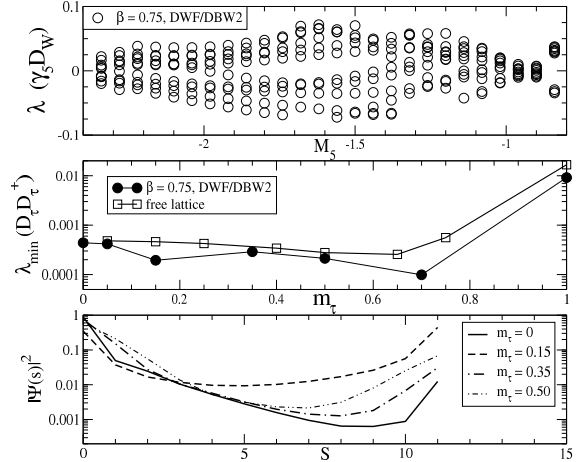}
\end{center}
\vspace{-1.4cm}
\caption{Up: Spectral flow of the test lattice. Middle: Lowest 
eigenvalue of $D_\tau D_\tau^+$ dependence on $m_\tau$ in the interacting and the free field case. 
Bottom: Eigenvector density distribution in the 5th dimension for different values of $m_\tau$.}
\label{fig:twist}
\end{figure}
We tested the effects of the twisted mass term on the eigenvalues 
and eigenmodes of $D_\tau D_\tau^+$, by running the Ritz algorithm on one chosen 
lattice from the dynamical DWF/DBW2 $\beta=0.75$ run. 
The graph in the middle of Figure~\ref{fig:twist} shows 
the dependence of the lowest eigenvalue of $D_\tau D_\tau^+$ on 
$m_\tau$. We see that for the range $m_\tau<0.8$, the eigenvalue is not strongly affected by 
the value of the twisted mass, a behavior very similar to the one observed 
in the free theory (shown on the same graph).
This is encouraging since it shows that $m_\tau$, 
as we hoped by its design, is an irrelevant parameter 
for a certain range of values, in the sense that it does not 
act as an additional fermion mass in the problem.
The dependence on $m_\tau$ of the density in the fifth dimension of 
the eigenvector corresponding to the lowest eigenvalue 
can be studied from the bottom graph on Figure~\ref{fig:twist}. Obviously the 
chirality and the localization of this mode 
changes with $m_\tau$ and we expect that this is true for the rest of the eigenmodes. 
Whether we can manipulate the system by
varying $m_\tau$ in a way that globally the zero-eigenmode condensation 
in the Aoki phase is suppressed, is still to be determined
and will be a subject of future investigations
\section{CONCLUSIONS}
The addition of an adjoint term to the DBW2 action in order to reduce further 
$m_{\rm res}$ at strong couplings did not give a significant improvement.
Increasing the amount of the rectangle term in the gauge action gives
an improvement in $m_{\rm res}$ at coarse scales, but at the possible expense 
of the distortion of short distance physics.
  
Finally, we showed that the twisted mass term in the DWF Dirac operator does affect the localization of the eigenvectors,
which we hope is an evidence that by tuning $m_\tau$ we can achieve 
better localization of the eigenstates and influence the onset of the Aoki phase.


\vspace*{-0.2cm}
\begin{thebibliography}{5} 
\bibitem{dbw2} T.~Takaishi, {\em Phys.Rev.} {\bf D 54}, 1050 (1996)
\bibitem{rbc} Y.~Aoki et al., {\em Phys.Rev.} {\bf D 69}, 074504 (2004)
\bibitem{inst} M.~Perez et al., {\em Nucl.Phys.} {\bf B 413}, 535 (1994)
\bibitem{me} L. Levkova et al., {\em Nucl.Phys.Proc.Suppl.} {\bf B 129\&130}, 399 (2004) 
\end{thebibliography}
\end{document}